\documentstyle[preprint,aps]{revtex}
\tighten

\begin{document}

\bibliographystyle{unsrt}

\preprint{NORDITA---93/62 A/revised}
\draft

%*********************   TITLE  *********************************

\title{
Negative energy densities in extended sources
generating closed timelike curves
in General Relativity
with and without torsion}

%********************   AUTHOR  *********************************

\author{Harald H. Soleng\footnote{
E-Mail address: HARALD@nordita.dk}}

\address{
NORDITA, Blegdamsvej 17,
DK-2100 Copenhagen {\O}, Denmark}

\date{November 1, 1993}

\maketitle

%*******************   ABSTRACT   *********************************

\begin{abstract}
Near a spinning point particle in (2+1)-dimensional gravity (or near an
infinitely thin, straight, spinning string in 3+1 dimensions)\
there is a region of space-time with closed timelike
curves. Exact solutions for extended sources with
apparently physically acceptable
energy-momentum tensors, have produced the same
exterior space-time structure. Here it is pointed out that in the case
with torsion, closed timelike curves appear only for spin
densities so high that the spin energy density is higher than the
net effective energy density. In models without torsion,
the presence of closed time-like curves is
related to a heat flow of unphysical magnitude. This
corroborates earlier arguments
against the possibility of closed timelike curves in
space-time geometries generated by physical sources.
\end{abstract}


%  \vskip2pc]

\renewcommand{\thefootnote}{\arabic{footnote}}
\addtocounter{footnote}{-\value{footnote}}

%%%%%%%%%%%%%%%%%%%%%%%%%%%%%%%%%%%%%%%%%%%%%%%%%%%%%%%%%%%%%%%%%%%

It has been conjectured
\cite{DeserJackiwtHooft}
that physically acceptable sources
together with reasonable boundary conditions
produce space-times
without causality violation
in (2+1)-dimensional gravity, and
in the case of spinless point particles,
it has been proved that
closed timelike curves (CTCs)\ cannot be realized
by physical timelike sources \cite{CFG,DJH,Kabat,Hooft}.
The appearance of CTCs
near a {\em spinning\/} point particle in
(2+1)-dimensional gravity \cite{DeserJackiwtHooft,GAK}
(or a spinning string in 3+1 dimensions \cite{Mazur})\
could be attributed to a
torsion singularity at the source \cite{Gerbert}.  Hence, in this case one
could argue that the causality violation is due to a
point (or line)\ singularity which by itself seems unphysical.

The singularity argument against these time-machines could be removed,
if one could find an extended source for this space-time without
losing the CTCs.
Within the
Einstein-Cartan theory \cite{Hehletal}, such a model
has been constructed \cite{Soleng}. It represents an infinite, straight
(3+1)-dimensional
string with spin-polarization along its axis. According to the
Einstein-Cartan
theory, the gravitational effect of spin is torsion. Although torsion does
not propagate in vacuum in this theory, the matching conditions at the
surface of a medium with torsion
lead to distant gravitational effects.
Thus, spin is known to produce exterior gravitational
fields similar to those induced by
orbital angular momentum \cite{Soleng2}.
The spinning string problem has been
studied within General Relativity without torsion, in which case
spin-polarization
is replaced
with orbital angular momentum \cite{JS}
(see also the related system in Ref.\ \cite{CCTC}), and it was claimed that
these constructions were examples of time-machines that respect the
energy conditions,
but according to a recent theorem by Menotti and Seminara \cite{MS}
a torsionless, stationary, and rotational symmetric
extended source would have to violate the weak energy
condition in order to produce CTCs. In this Brief Report this
contradiction is resolved by a careful reexamination of
the models in
Refs.\ \cite{Soleng,JS,CCTC}, and it
will be shown that
the sources are indeed unphysical when CTCs exist. Thus, the present study
corroborates the earlier results of Refs.\ \cite{CFG,DJH,Kabat,Hooft,MS}
for Einstein's theory and
indicates an extension to the Einstein-Cartan theory.

The geometry of the model is specified by the orthonormal
one-forms (tetrad frame)
\begin{equation}
\left.\begin{array}{ccl}
\omega^{0}&=& dt+Md\phi\\
\omega^{1}&=& dr \\
\omega^{2}&=&\rho d\phi\\
\omega^{3}&=&dz .
\end{array} \right.
\end{equation}
$M$ and $\rho$ are functions of $r$. The metric in the coordinate frame
is found by using that $ds^2=\eta_{\mu\nu}\omega^{\mu}\omega^{\nu}$.
When torsion is polarized along
the $z$-axis, we get only one nonvanishing component of the torsion tensor,
namely
\begin{equation}
T^{0}_{\;\;12}=\sigma.
\end{equation}
Defining
\begin{equation}
\Omega\equiv -\frac{1}{2}\sigma+\frac{M'}{2\rho},
\label{omdef}
\end{equation}
the Einstein tensor takes the form \cite{Soleng}
\begin{equation}
\left.\begin{array}{ccl}
G^{0}_{\;\;0}&=&-3\Omega^2-\sigma\Omega+\rho''/\rho\\
G^{1}_{\;\;1}&=&
G^{2}_{\;\;2}=\Omega^2\\
G^{3}_{\;\;3}&=&-\Omega^{2}-\sigma\Omega+\rho''/\rho\\
G^{0}_{\;\;2}&=&-\Omega'.
\end{array} \right.  \label{Einstein}
\end{equation}
If
\begin{equation}
\rho=\frac{\sin(\sqrt{\lambda}r)}{\sqrt{\lambda}} ,
\label{rhor}
\end{equation}
and torsion is
nonzero and constant throughout the source, one can integrate the field
equations assuming that both the heat flow, $G^{0}_{\;\;2}$,
and the radial pressure, $G^{1}_{\;\;1}$,  vanish. Then using
units such that $8\pi G =1$, the
energy-momentum tensor takes the form $T_{\mu\nu}=
{\mbox{diag}}[\lambda,0,0,-\lambda]$ which satisfies the weak
energy condition by construction.

This model can be matched to a torsionless vacuum characterized by the
orthonormal frame
\begin{equation}
\left.\begin{array}{ccl}
\theta^{0}&=& dt+\frac{j}{2\pi}d\phi\\
\theta^{1}&=& dr\\
\theta^{2}&=& \left(1-\frac{\mu}{2\pi}\right)\left(r+r_{0}\right)d\phi\\
\theta^{3}&=& dz .
\end{array} \right.
\end{equation}
The parameters in the exterior metric are determined by the
Arkuszewski-Kopczy{\'n}ski-Ponomariev junction conditions \cite{AKP}
appropriate for the Einstein-Cartan theory. The result is (see Ref.\
 \cite{Soleng} for details)
\begin{equation}
\left.\begin{array}{ccl}
j&=& M(R)\\
\mu&=&2\pi \left(1-\cos(\sqrt{\lambda}R)\right)\\
\sqrt{\lambda}r_{0}&=&\tan(\sqrt{\lambda}R)-\sqrt{\lambda}R.
\end{array} \right.  \label{jmur}
\end{equation}
Here $R$ is the surface radius of the spinning string.
$\mu$ is the angle deficit in the exterior geometry, and in the
present units, it is also the mass per length of the string. $j$ is the
angular momentum per length.
{For} realistic strings, with $\mu\ll 2\pi$, CTCs can exist only for
unrealistic spin densities \cite{Soleng}. In
general,
closed timelike curves appear at
the surface of the string provided $M(R)^2>\rho(R)^2$ because $\phi$ is then
a timelike coordinate, By use of Eqs.\ (\ref{rhor})
and (\ref{jmur}), this condition can be written
\begin{equation}
\left(1-\frac{\mu}{2\pi}\right)^2> 1-\left(\frac{j}{2\pi}\right)^{2}\lambda ,
\end{equation}
which due to the relation $j=\sigma/\lambda \mu$, takes the simple form
\begin{equation}
\mu > 4\pi\left(1+\sigma^2/\lambda\right)^{-1}.
\end{equation}
But $\mu<2\pi$, in a conical universe.
This leads to the requirement that
\begin{equation}
\sigma^{2}>\lambda .
\end{equation}
Thus, in order that the spinning string in the Einstein-Cartan model
induce a causality violating space-time, the spin energy density
$\sigma^2$ has to be larger than the net effective energy density, $\lambda$.
If the effective energy density $\lambda$ is a sum of the spin energy and
a `bare' energy, it follows that this `bare' energy density has to be
negative.

In the case of vanishing torsion, it is impossible to avoid a
heat flow. To generate CTCs, the source has to produce
a nonvanishing $M(r)$, and to avoid a torsion singularity at the origin,
$M(0)=0$. It follows that
$M(r)$ is not a constant, and by the definition (\ref{omdef}),
that and $\Omega(r)\neq 0$. Now, to match smoothly to
the exterior vacuum, the radial pressure at the surface has to vanish.
Since by the form of the Einstein tensor (\ref{Einstein}),
$p_{r}=G^{1}_{\;\; 1}=\Omega^2$, vanishing surface pressure
implies that $\Omega (R)=0$.
Because $\Omega$ must be zero at one point and nonzero elsewhere,
$\Omega'(r)\neq 0$. In Refs.\ \cite{JS,CCTC}, it was
argued  that by taking $\rho (r)$ as the same function as in the
torsion case,
cf.\ Eq.\ (\ref{rhor}),
the energy density becomes positive definite by construction. Then $\Omega'$
was taken to be a constant of magnitude equal or less than the minimal
energy density of the model. It then followed from Einstein's field equations
that
\begin{equation}
T^{00}\geq |T^{\mu i}|,
\end{equation}
and
it was assumed that the dominant energy condition was satisfied.
This is wrong. Indeed, consider an energy-momentum tensor
of the form
\begin{equation}
T=\left[\begin{array}{cc}
         \rho  & q \\
          q    & p
        \end{array}
  \right] ,
\end{equation}
where $\rho$ is the energy
density, $q$ is the heat flow, and $p$ is the pressure, and where
the spatial dimensions orthogonal to the heat flow have been supressed.
Boost
it by use of the Lorentz transformation
\begin{equation}
L=\left[\begin{array}{cc}
          \gamma  & -u\gamma\\
          -u\gamma & \gamma
        \end{array}
  \right].
\end{equation}
Then one finds (in the limit $u=1$)\ that the energy density in
the new frame is given as
\begin{equation}
\rho'=(\rho-2q+p)\gamma^2.
\end{equation}
Hence, the weak energy condition is violated unless the magnitude of the
heat flow, $q$, is everywhere equal or less than half the sum of
energy density and pressure. Thus, the time machines of Refs.\ \cite{JS,CCTC}
require an energy-momentum tensor that corresponds to a negative
energy density in some frames, and these solutions are therefore not
in conflict with the general result that the energy conditions protect
against CTCs \cite{MS}.

I thank Professor S.~Deser for bringing the paper of
Menotti and Seminara to my attention and for comments on
time machines in (2+1)-dimensional gravity.
This work was supported in part by
the Thomas Fearnley Foundation and  by
Lise and Arnfinn Heje's Foundation, grant number 0F0377/1993.

%*******************      REFERENCES    *************************

\end{document}